\documentclass[sigconf,nonacm]{acmart}

\usepackage[utf8]{inputenc}
\usepackage[table]{xcolor} 
\usepackage{colortbl}
\usepackage{subcaption}
\usepackage{graphicx}
\usepackage{multirow}
\usepackage{booktabs}
\usepackage{tabularx}
\usepackage{makecell}
\usepackage{amsmath}   
\usepackage{amsfonts}  
\usepackage[dvipsnames,table,svgnames]{xcolor}
\usepackage[framemethod=TikZ]{mdframed}
\usepackage{enumitem}
\usepackage{longtable}
\usepackage{tabularx}
\usepackage{multirow}
\usepackage{supertabular}
\usepackage{hyperref}
\usepackage{array}
\usepackage{listings}
\usepackage{stfloats}
\usepackage{cuted}

\usepackage{amssymb}   
\usepackage{bm}        
\usepackage{dsfont}    
\usepackage{pifont}    

\usepackage{tikz}

\usepackage{adjustbox}
\usetikzlibrary{positioning}

\definecolor{rootcol}{RGB}{45,45,45}
\definecolor{catredcol}{RGB}{160,30,30}
\definecolor{catbluecol}{RGB}{20,70,150}
\definecolor{errredbg}{RGB}{255,235,235}
\definecolor{errbluebg}{RGB}{232,240,255}

\tikzset{
  base/.style={draw, rounded corners=3pt, align=left, inner sep=6pt, line width=0.7pt},
  rootNode/.style={base, fill=rootcol, text=white, font=\bfseries\large,
                   align=center, rounded corners=5pt},
  catRed/.style={base, fill=catredcol, text=white, font=\bfseries\large,
                  align=center, rounded corners=4pt},
  catBlue/.style={base, fill=catbluecol, text=white, font=\bfseries\large,
                   align=center, rounded corners=4pt},
  errBox/.style={base, text=black, font=\large,
                 minimum width=4.6cm, minimum height=2.6cm, text width=4.4cm},
  errRed/.style={errBox, fill=errredbg, draw=catredcol!60},
  errBlue/.style={errBox, fill=errbluebg, draw=catbluecol!50},
  myedge/.style={draw=gray!70, line width=0.65pt},
}

\newcommand*\circled[1]{\tikz[baseline=(char.base)]{
    \node[shape=circle,draw,inner sep=1pt, font=\small\bfseries, fill=white, text=black] (char) {#1};}}

\definecolor{lightgray}{gray}{0.95}
\definecolor{PastelGreen}{RGB}{232, 250, 235}

\newmdenv[
  linecolor=black!30,
  linewidth=0.8pt,
  roundcorner=4pt,
  innertopmargin=4pt,    
  innerbottommargin=4pt, 
  innerleftmargin=4pt,   
  innerrightmargin=4pt,   
  backgroundcolor=PastelGreen, 
  skipabove=4pt,           
  skipbelow=0pt,           
]{emphasizedBox}

\newmdenv[
  linecolor=black!30,
  linewidth=0.8pt,
  roundcorner=4pt,
  innertopmargin=4pt,    
  innerbottommargin=4pt, 
  innerleftmargin=4pt,   
  innerrightmargin=4pt,   
  backgroundcolor=lightgray, 
  skipabove=4pt,           
  skipbelow=0pt,           
]{insightBox}

\AtBeginDocument{%
  }

\begin{document}

\title{Both Ends Count!\\Just How Good are LLM Agents at Text-to-``Big SQL''?}
\titlenote{Accepted at EuroMLSys 2026}


\author{Germán T. Eizaguirre}
\affiliation{%
  \institution{Universitat Rovira i Virgili}
  \city{Tarragona}
  \country{Spain}}
\email{germantelmo.eizaguirre@urv.cat}

\author{Lars Tissen}
\affiliation{%
  \institution{RWTH Aachen University}
  \city{Aachen}
  \country{Germany}
}
\email{lars.tissen@rwth-aachen.de}

\author{Marc Sánchez-Artigas}
\affiliation{%
  \institution{Universitat Rovira i Virgili}
  \city{Tarragona}
  \country{Spain}
}
\email{marc.sanchez@urv.cat}

\renewcommand{\shortauthors}{Eizaguirre et al.}

\begin{abstract}
Text-to-SQL and Big Data are both extensively benchmarked fields, yet there is limited research that evaluates them jointly. In the real world, Text-to-SQL systems are often embedded with Big Data workflows, such as large-scale data processing or interactive data analytics. We refer to this as \textbf{``Text-to-Big SQL''}. However, existing text-to-SQL benchmarks remain narrowly scoped and overlook the cost and performance implications that arise at scale. For instance, translation errors that are minor on small datasets lead to substantial cost and latency overheads as data scales, a relevant issue completely ignored by text-to-SQL metrics.

In this paper, we overcome this overlooked challenge by introducing novel and representative metrics for evaluating Text-to-Big SQL. Our study focuses on production-level LLM agents, a database-agnostic system adaptable to diverse user needs. Via an extensive evaluation of frontier models, we show that text-to-SQL metrics are insufficient for Big Data. In contrast, our proposed text-to-Big SQL metrics accurately reflect execution efficiency, cost, and the impact of data scale. For example, GPT-4o compensates for roughly 7\% lower accuracy than the top-performing later-generation models with up to a 12.16× speedup, while GPT-5.2 is more than twice as cost-effective as Gemini 3 Pro at large input scales.

\end{abstract}


\maketitle

\section{Introduction}

Text-to-SQL is a longstanding problem in NLP that seeks to bridge natural language (NL) interfaces and structured query generation. Recent advances in production Large Language Models (LLMs) have substantially improved cross-domain performance, placing them as effective text-to-SQL engines
\cite{anthropic2024claude3opus,openai2025gpt52,chung2025long-context} that can generalize across varying data schemas and terminology, improving state-of-the-art results~\cite{zhu2024largelanguagemodelenhanced,li2024dawn}.

In practice,
this generalization ability is further enhanced via the integration of AI agents, which act as task-specific LLM scaffolds to iteratively inspect database schemas, refine SQL generation, validate SQL  syntax,  enabling text-to-SQL
execution to adapt to the unique traits of user-specific data sources~\cite{sapkota2026agents}.
Precisely, the existing ecosystem of open-source stacks for agent implementation \cite{microsoft2024autogen,langchain2026,crewai2026home}, combined with the current state of production LLMs \cite{anthropic2024claude3opus,openai2025gpt52,chung2025long-context}, has made text-to-SQL systems more accessible than ever.

However, when moving beyond traditional databases to Big Data systems, text-to-SQL faces additional complexities. For instance, systems such as Amazon Athena~\cite{aws_athena_overview}
enable interactive analytics on massive datasets without requiring ETL, supporting various formats (CSV, JSON, ORC, Avro) and serverless, on-demand execution. In Big Data systems such as Athena, focusing only on the text-to-SQL end is not enough. The Big Data end itself
introduces critical constraints that directly affect overall performance and cost. First,
incorrect SQL has amplified consequences: failed queries can consume substantial compute resources, scan massive volumes of data, and increase execution costs, making accuracy essential not only for correctness but also for efficiency. For instance, \cite{KoutsoukosM0K25}
report that running TPC-H benchmark at a moderate scale factor of $100$ on Amazon Athena using Parquet data takes $132.3$ seconds, illustrating how failed queries can sharply increase execution time and costs in Big Data environments.

Second, inefficiencies can arise not only from failed queries but also from the SQL generation process itself. In agentic text-to-SQL, the LLM instructs the agent to use structured tools to inspect schemas and extract data-specific traits~\cite{deng2025reforcetexttosqlagentselfrefinement,xie2024magsqlmultiagentgenerativeapproach,zhang2024sqlfuseenhancingtexttosqlperformance}. While this enables context-aware query adaptation, the reasoning overhead and tool orchestration can increase latency, making efficient SQL generation critical in Big Data systems. Simply put,
if SQL generation becomes slower than physical query execution, interactive analysis may become impractical, undermining the performance gains of decades of optimizations in Query-as-a-Service (QaaS) engines like Athena and BigQuery~\cite{google_cloud_bigquery}.

Overall, these reasons lead to the following observation:

\begin{emphasizedBox}
Considering both SQL generation and execution is crucial, making text-to-SQL in Big Data environments distinct, for we refer to it as ``\textbf{Text-to-Big SQL.}''\footnotemark
\end{emphasizedBox}
~\footnotetext{Here, ``Big SQL'' denotes a conceptual domain in SQL query processing; it is unrelated to IBM's Db2 Big SQL product~\cite{ibm_db2_big_sql}. ``Big SQL'' is a trademark of IBM Corporation, used descriptively and with no affiliation.}

\subsection{Why Current Text-to-SQL Approaches Fall Short for Big Data?}

Traditional text-to-SQL methods have largely been evaluated on moderate-scale relational databases,
focusing on query translation or isolated accuracy metrics~\cite{spider202025,li2024dawn,zhang2024finsql,deochake2025costawaretexttosqlempiricalstudy}.
While these benchmarks provide insights into LLM capabilities, they often overlook the complexities of interactive execution,
streaming data, and cost considerations~\cite{cheng2025barbariansgateaiupending}.

One example of this is that many text-to-SQL benchmarks use binary correctness metrics, tagging each generated query with a simple $0$/$1$ label,  thereby diluting degrees of partial correctness~\cite{li2023bird,zhang2024finsql}. This is obvious, for example, when a generated query incorrectly projects an unnecessary column. In traditional text-to-SQL this counts as a wrong translation, but in text-to-Big SQL it should be partially acceptable, since re-running a query for a single extra column could be extremely costly. Consequently, new evaluation metrics are needed to jointly account for  partial correctness and cost, reflecting practical text-to-SQL performance for Big Data.

While LLMs are effective text-to-SQL processors~\cite{li2025OmniSQL,li2024dawn}, their performance depends on how agents scaffold tool use~\cite{sapkota2026agents,yao2023react,liu2025supportingaioverlordsredesigning}. In a ReAct-style framework, the LLM controller guides reasoning, selects tools, and interprets feedback, while the executor runs the tools. Fast tools, such as fetching a table schema from a data catalog can be bottlenecked by extensive LLM reasoning for query validation, or vice versa. Efficient interaction between LLM, agent, and tools is thus critical for responsive interactive analytics. However, there are
no evaluations that focus on this interplay and its effect on Big Data performance. In this paper, we start addressing this gap by jointly measuring agent and tool utilization, reasoning latency, and downstream query execution.

\subsection{Our Contribution}

In this work, we propose a benchmarking methodology for text-to-Big SQL agents that treats both ends, namely query generation and execution, as first-class citizens. We focus on \textbf{zero-shot} LLM agents to
examine a worst-case scenario where no specific fine-tuning or additional optimization is applied,
revealing the true impact of SQL generation, agent action,
and tool interaction on performance
and cost in Big Data settings. Our contributions are the following:

\begin{enumerate}
    \item A novel evaluation framework for text-to-SQL agents designed to capture big query execution.
     We propose new metrics that jointly assess agent action, reasoning latency, and the cost-effectiveness
     of generated queries, reflecting both partial correctness
      and the practical implications of running queries on Big Data engines.
    \item A systematic evalution of state-of-the-art LLMs within a unified ReAct-style agent architecture. This analysis reveals insights
     beyond accuracy, identifying scenarios where newer models achieve high correctness but are less interactive due to reasoning or
     tool orchestration overhead, which happens for instance with Opus 4.6.
    \item A discussion of the unique challenges, as well as open research questions in text-to-Big SQL, including the interplay
     between SQL generation, agent tool use, and execution performance, an area largely overlooked by existing benchmarks.
\end{enumerate}

\section{Text-to-Big SQL Demands New Metrics}
\label{sec:metrics}


\subsection{Limitations of Current Metrics}

Typically, a text-to-SQL benchmark suite comprises a set of triples containing a\textit{ natural language} (NL) query, a \textit{golden query} in SQL, and a \textit{ground truth} ($V^n$) result~\cite{li2023bird}. During evaluation,  system performance is measured by comparing the generated SQL to the golden query, and the resulting output ($\hat{V}^n$) to $V^n$. Overall accuracy is then computed by aggregating these comparisons across the benchmark suite using standard evaluation metrics, such as Exact Matching (EM), Execution Accuracy (EA), and Valid Efficiency Score (VES)~\cite{hong2025database-interfaces,zhu2024largelanguagemodelenhanced,luo2025state-of-the-art}.

The major limitation of these metrics is their reliance on all-or-nothing correctness\footnote{The corresponding formulas are provided in the Appendix.}. In practice, however, a generated query may deviate from the expected result without being entirely invalid, provided that
end users may still be able to determine its correctness through simple validation, for example, by detecting an additional column in the projected output. We summarize the possible outcomes of an SQL execution as follows:

\noindent \circled{A} \textbf{Incorrect row count}: The result is invalid due to poor SQL translation, for example, mis-specified \texttt{WHERE} conditions or inappropriate join types (\texttt{INNER}, \texttt{LEFT}, etc.), leading to an incorrect number of rows. Such silent failures may remain unnoticed by the user, but always need a new translation and re-execution cycle to ensure correctness.

\noindent \circled{B} \textbf{Missing columns}: The result is invalid due to omission of mandatory attributes in the
projection list, requiring query re-execution to retrieve the complete output by adding the missing attributes in the \texttt{\small SELECT} clause.

\noindent \circled{C} \textbf{Superfluous columns}: The result is valid, as we assume that experienced users can manually drop the extra columns without modifying the returned output, which is very cheap and fast. For instance, \texttt{\small df.drop(``extra\_col'')} returns a new DataFrame without the extra column in Spark~\cite{ApacheSparkSQL}. However, processing superfluous data affects execution performance and cost and should be penalized in the assessment.



\subsection{Proposed Metrics}

To encode our notion of query correctness in a measurable metric,
we extend the standard text-to-SQL VES metric to account for superfluous columns. Specifically, to quantify the overhead introduced by including irrelevant columns, we compute the column-level precision: $
P(S, \hat{S}) = \frac{|S \cap \hat{S}|}{|\hat{S}|}$,
where $S$ is the set of ground-truth columns and $\hat{S}$ is the~set of columns in the execution result. This captures the fraction of retrieved columns that are actually relevant, penalizing additional, unnecessary columns without discarding partially correct results (\circled{C}).

Also,
our new metric considers the total end-to-end (e2e) time ($T_{\text{e2e}}$), which includes all back-and-forth interactions between the LLM and the agent, the execution of interim tools, as well as the time required to run the generated SQL query on the underlying Big Data engine. Combining both concepts, we propose the novel text-to-Big SQL metric called $\mathrm{VES}^*$, which is defined as follows for $N$ queries:

\begin{equation} \label{eq:ves}
\mathrm{VES}^* = \frac{1}{N} \sum_{i=1}^{N} \left( \mathds{1}(V_i, \hat{V}_i) \cdot P(S_i,\hat{S}_i) \cdot \frac{T_{gold}}{T_{e2e}} \right),
\end{equation}
where $T_{gold}$ denotes the execution time of the golden query, and $\mathds{1}(V_i, \hat{V}_i)$ is an indicator function that tells whether the output of the generated query $\hat{V}_i$ matches the expected result $V_i$ (\circled{A} and \circled{B}):

\begin{equation}
\mathds{1}(V, \hat{V}) =
\begin{cases}
1, & \text{if $V$ is contained in expected output $\hat{V}$}\\
0, & \text{otherwise}
\end{cases},
\label{eq:accuracy-indicator}
\end{equation}

We also introduce the \textit{Valid Cost-Efficiency Score} (VCES), a cost-oriented derivative of $\text{VES}^*$ to account for the overall execution cost $C_{\text{e2e}}$, including the iterative interactions between the LLM and agent, the execution of agent-invoked tools, and the runtime of the generated query on the Big Data engine. For text-to-Big SQL, benchmarking query execution cost is relevant, particularly in cloud deployments:

\begin{equation} \label{eq:vces}
\text{VCES} = \frac{1}{N} \sum_{i=1}^{N} \left( \frac{\mathbf{1}(V_i, \hat{V}_i)  \cdot P(S_i,\hat{S}_i) \cdot \frac{T_{gold}}{T_{e2e}}}{C_{e2e}} \right).
\end{equation}

To complement the previous metrics, we introduce the \textit{Expected Cost per Valid Query} (CVQ), which quantifies the anticipated cost to obtain a valid result under a retry-until-success strategy. Let $p$ denote the single-shot validity rate, that is, the fraction of generated big queries that are valid. Consequently, the expected number of attempts until success follows a geometric distribution with mean $1/p$. Accordingly, we define $CVQ = \frac{C_{\text{e2e}}}{p}$.

\section{AI Agent Design}

\subsection{Decision-Making Logic}

To evaluate the novel text-to-Big SQL paradigm, we use a ReAct (Reasoning + Acting) agent~\cite{yao2023react}, a well-established framework~\cite{sapkota2026agents, paperbench2025,liu2025palimpchat}, which decomposes agentic operation into three intertwined components: \textit{Thought}, \textit{Action}, and \textit{Observation}. Via \textit{Thought}, the agent reasons about the task and decides its next step; through \textit{Action}, it interacts with the environment or an external tool to gather or process information; and through \textit{Observation}, it interprets feedback from these actions to refine subsequent reasoning steps.

We deliberately keep the agent simple to provide a broader perspective in our analysis. While more complex AI agents exist~\cite{li2025alphasqlzeroshottexttosqlusing}, we opt to leverage the long-context capabilities of production LLMs guided by iterative cycles of Thought, Action, and Observation, which has already demonstrated its effectiveness in text-to-SQL tasks~\cite{chung2025long-context,liu2025supportingaioverlordsredesigning}.

We define the \textit{controller} as the connected LLM that guides ReAct-style reasoning, decides when and how to use tools, and produces the final answer. Conversely, the \textit{executor} is the program that connects the LLM with external tools and handles the execution loop.

\subsection{Agentic Architecture}

\begin{figure*}[htbp]
    \centering
    \includegraphics[width=0.9\textwidth]{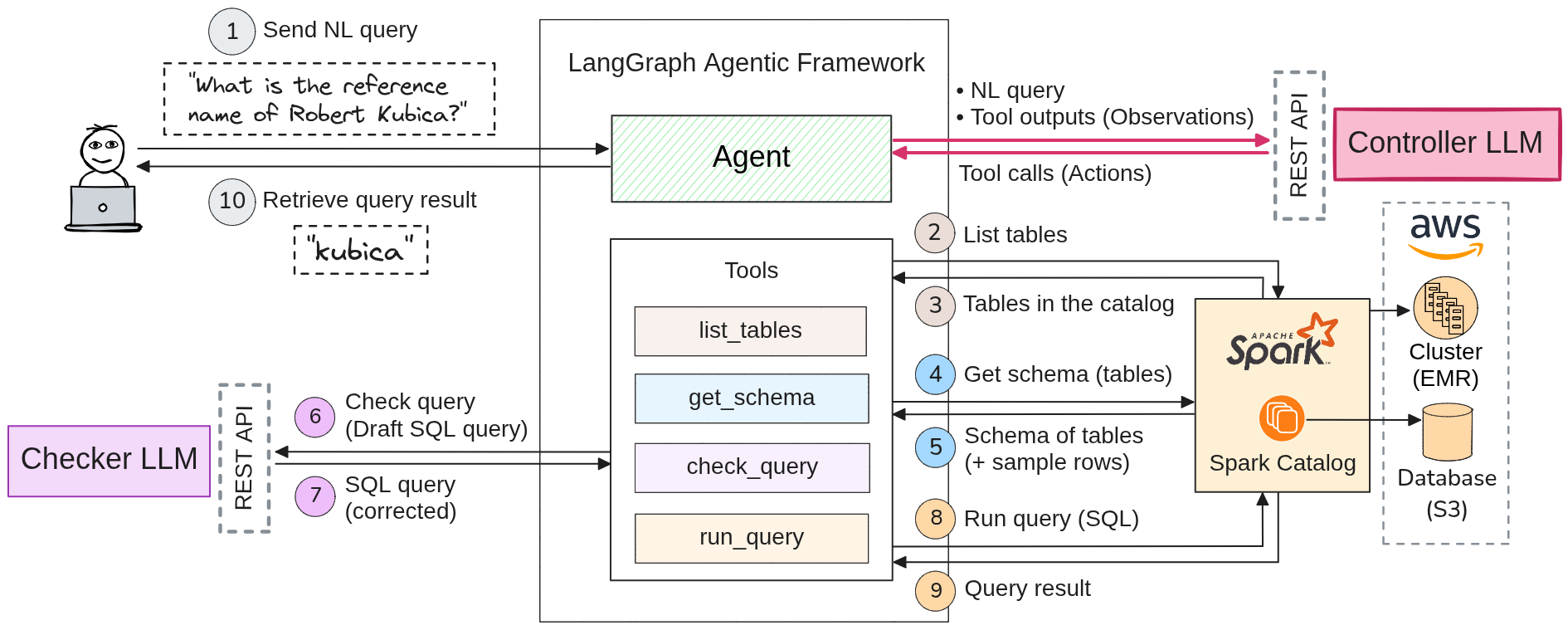}
    \caption{Architecture and typical execution flow of the evaluated LLM agent.}
    \label{fig:agent-arch}
\end{figure*}

The architecture of our agent is illustrated in Figure~\ref{fig:agent-arch}. For the downstream query engine, we choose Spark SQL~\cite{ApacheSparkSQL} due to its widespread adoption for large-scale structured data analysis. Although this work focuses on Spark, the evaluated agentic workflow is compatible with most data analytics frameworks and backend deployments. The Spark session can be connected to either a local deployment or a distributed cluster.

The controller interacts with the Spark session through a set of four tools, whose specifications are fully inserted into the context in a zero-shot manner \cite{zero-shot-tools2023}. 
We define the tools as follows:

\begin{description}[leftmargin=1.5em, labelsep=0.5em]
    \item[\circled{1}] \texttt{list\_tables}: Look up the  Spark Catalog~\cite{pyspark_catalog_docs} to retrieve available tables by executing a \texttt{SHOW TABLES} statement.
    \item[\circled{2}] \texttt{get\_schema}: Retrieve the schema for one or more tables from the Spark Catalog using \texttt{SHOW CREATE TABLE t}. Optionally, the controller can fetch an adjustable number of sample rows from each table with the same tool call, which triggers an additional \texttt{SELECT * FROM t} statement.
    \item[\circled{3}] \texttt{check\_query}: Verify the syntax of a proposed query using predefined heuristics~\cite{chung2025long-context} using a connected LLM (the ~\textit{checker}).
    \item[\circled{4}] \texttt{run\_query}: Execute a SQL query in the connected Spark Session, regardless of whether the deployment is local or cluster-based.
\end{description}

We base our implementation on the original LangChain Spark SQL Toolkit~\cite{langchain_sparksql_2026} but port it to LangGraph~\cite{langchain_langgraph_2024}. 
We use the LangChain stack because it is actively maintained~\cite{langchain2026} and frequently serves as a research baseline~\cite{ding2025unifiedtoolintegrationllms, bhagat-etal-2025-evaluating, heng2025langchain-hpc}.

The proposed agent could be adapted for more fine-grained component optimization or more flexible workflows. To wit, both our controller and checker operate on the same LLM, and the same controller is leveraged for all iterations of the ReAct framework. Instead, independent LLMs could be tuned for performance or cost-efficiency at the iteration level (see Section~\ref{sec:discussion}).

\subsection{Text-to-Big SQL Workflow}

Listing~\ref{lst:agent-trace} provides an example of the typical execution flow of the proposed agent. The controller begins by listing all tables available within the connected database using the \texttt{list\_tables} tool (steps \ding{173}-\ding{174}). Although the agent currently supports a single connection, it can be extended to a multi-database catalog similar to other benchmarks~\cite{spider202025}, a feature Spark supports natively~\footnote{\url{https://spark.apache.org/docs/latest/sql-data-sources-jdbc.html}}. Next, the controller  identifies the involved tables in the query and retrieves their attributes and data types from the catalog calling \texttt{get\_schema}. This tool returns schema to the controller in a series of symbolic \texttt{CREATE TABLE}s. To further guide query construction the controller may optionally sample a custom number of rows as a practical hint  (steps \ding{175}-\ding{176}).
Once the controller decides that it has gathered sufficient database metadata, it generates a SQL query and dispatches it to the checker for necessary syntax corrections (steps \ding{177}-\ding{178}). The final SQL query is then executed within the Spark Session by calling the \texttt{run\_query} tool (step \ding{179}), with either the resulting data or the execution error returned to the user (step \ding{180}).


\definecolor{thoughttext}{RGB}{120,80,150}
\definecolor{actiontext}{RGB}{30,80,140}
\definecolor{obstext}{RGB}{120,100,60}
\definecolor{resulttext}{RGB}{40,120,60}

\begin{lstlisting}[
  language=bash,
  caption={Typical execution trace of the Text-to-SQL agent. Although the ReAct paradigm explicitly models reasoning as a ‘Thought’ step, in modern LLMs reasoning may occur internally and thus may not directly observable in traces. Instead, LLMs may return tool calls directly~\cite{su2026brokenchainscostincomplete}.},
  label={lst:agent-trace},
  basicstyle=\ttfamily\footnotesize,
  frame=none,
  backgroundcolor=\color{gray!5},
  numbers=none,
  aboveskip=5pt,
  belowskip=5pt,
  showstringspaces=false,
  breaklines=true,
  breakatwhitespace=true,
  columns=flexible,
  keepspaces=true,
  escapeinside={(*@}{@*)}
]
(*@\textbf{\textcolor{black!70}{Input NL Query:}}@*) "How many superheroes have blue eyes?"

(*@\textbf{\textcolor{thoughttext}{Thought:}}@*) I need to first check what tables are available...
(*@\textbf{\textcolor{actiontext}{Action:}}@*) list_tables_sql_db
(*@\textbf{\textcolor{actiontext}{Action Input:}}@*) {}
(*@\textbf{\textcolor{obstext}{Observation:}}@*) superhero, superpower, alignment...

(*@\textbf{\textcolor{thoughttext}{Thought:}}@*) I should check the schema of the tables...
(*@\textbf{\textcolor{actiontext}{Action:}}@*) schema_sql_db
(*@\textbf{\textcolor{actiontext}{Action Input:}}@*) {'table_names': 'superhero, colour'}
(*@\textbf{\textcolor{obstext}{Observation:}}@*) CREATE TABLE colour (id int, colour string);
CREATE TABLE superhero (id int, superhero_name string, ...);

(*@\textbf{\textcolor{thoughttext}{Thought:}}@*) Let me write the query and have it checked first.
(*@\textbf{\textcolor{actiontext}{Action:}}@*) query_checker_sql_db
(*@\textbf{\textcolor{actiontext}{Action Input:}}@*) {'query': "SELECT COUNT(s.id) FROM superhero s
JOIN colour c ON s.eye_colour_id = c.id WHERE c.colour = 'Blue'"}
(*@\textbf{\textcolor{obstext}{Observation:}}@*) SELECT COUNT(`s`.`id`) FROM `superhero` `s` JOIN `colour` `c` ON `s`.`eye_colour_id` = `c`.`id` WHERE `c`.`colour` = 'Blue'

(*@\textbf{\textcolor{thoughttext}{Thought:}}@*) The query is correct, so I can execute it.
(*@\textbf{\textcolor{actiontext}{Action:}}@*) query_sql_db
(*@\textbf{\textcolor{actiontext}{Action Input:}}@*) {'query': "SELECT COUNT(`s`.`id`) FROM `superhero` `s` JOIN `colour` `c` ON `s`.`eye_colour_id` = `c`.`id` WHERE `c`.`colour` = 'Blue'"}
(*@\textbf{\textcolor{obstext}{Observation:}}@*) [('234',)]

(*@\textbf{\textcolor{thoughttext}{Thought:}}@*) I now know the final answer.
(*@\textbf{\textcolor{resulttext}{Final Answer:}}@*) (*@\textbf{234}@*)

\end{lstlisting}

To prevent the controller from iteratively correcting and re-running queries, we terminate the agent immediately following the first \texttt{run\_query} execution. We take this design decision because, in Big Data systems, unrestricted execution loops can lead to excessive resource consumption or high billing costs, and may even result in \textit{stuck-in-the-loop}~\cite{cheng2025barbariansgateaiupending} scenarios without improving the accuracy of the inferred query.

While this trace reflects the most frequent tool-call sequence, the actual execution flow may vary depending on the controller LLM, which may repeat or omit specific tools.

\section{Evaluation}
\label{sec:evaluation}

Our evaluation proceeds as follows. \S\ref{sec:evaluation-tools} shows that text-to-SQL metrics lack informativeness, revealing opportunities from fine-grained agent benchmarking. \S\ref{sec:evaluation-metrics} demonstrates the superior discriminability of text-to-Big SQL versus text-to-SQL metrics in differentiating agents across latency and cost objectives. Finally, \S\ref{sec:evaluation-scale} shows data scale matters and that its impact can be quantified via text-to-Big SQL metrics. We employ two benchmarks: a text-to-SQL-focused one and a Big Data-focused one.

\begin{itemize}
    \item \textbf{BIRD~\cite{li2023bird}}, a text-to-SQL benchmark that assesses translation accuracy for realistic databases.
    \item \textbf{TPC-H~\cite{tpc_h_spec}}, a classic data analytics benchmark that measures database performance on complex ad-hoc business queries over relational data. TPC-H is very useful for text-to-Big SQL because it allows for the deterministic scaling of data.
\end{itemize}

We conducted all experiments on an AWS \texttt{m5.xlarge} EC2~\cite{aws_ec2_product} instance in the \texttt{us-east-1} region. We selected a set of representative frontier models from various providers for the evaluation, according to the following criteria: (1) current frontier models from Google, Anthropic, and OpenAI; (2) previous-generation frontier models from these same providers; and (3) the three open-source models with the highest scores on SWE-rebench~\footnote{\url{https://swe-rebench.com/}} (all at the time of the experiment, mid-February 2026)~\footnote{Detailed results for each model are available in the Appendix.}. For all models, we used the official provider APIs and calculated costs based on their respective per-token pricing.

To ensure consistency for interactive use, all models are deployed with low-latency reasoning configurations. We standardize sampling hyperparameters, such as temperature, top-$p$, and maximum token limits, across all APIs, except where specific parameters are not exposed by the provider.

\subsection{Accuracy is Not Enough in Text-to-Big SQL}
\label{sec:evaluation-tools}

When models achieve similar accuracy, standard text-to-SQL metrics fail to differentiate between setups.
We prove this by inspecting the Execution-based Focused Evaluation (EX)~\cite{spider202025}, a SOTA text-to-SQL accuracy metric. Since text-to-SQL performance continues to improve independently~\cite{hong2025database-interfaces}, we design a practical testbed assuming near-perfect accuracy.  To this end, we select eight queries from the BIRD dataset where all tested models (except the later-added GPT-5.2) attain an average EX of at least 0.85.

\textbf{EX and e2e execution time alone lack the informativeness to discriminate models effectively.} Figure~\ref{tab:time-perc-breakdown-combined} shows this: for faster models, the accuracy/speed tradeoff is unclear: e.g., is GPT-4o better than Gemini 3 Flash (27.79\% faster but imperfect accuracy)? \textbf{If wrong-query costs are high, a slightly slower but more accurate model may be preferable in production}.

Later-generation models do not clearly outperform their predecessors in zero-shot agentic text-to-SQL. For example, Opus 4.6 achieves perfect accuracy but takes 92.37\% longer execution time than GPT-4o. Similarly, Gemini 3 Pro and GLM-5 exhibit poor latency, further exacerbated by high variance from API instabilities. Notably, GPT-4o was released in 2024, nearly two years before the other two models.

\begin{table}[ht]
\centering
\caption{EX, total execution time and the breakdown of execution time per "Observation-Thought-Action" stage for each LLM, on selected BIRD queries. Results are displayed in ascending order of total execution time.}
\label{tab:time-perc-breakdown-combined}
\setlength{\tabcolsep}{2.5pt}
\footnotesize
\begin{tabularx}{\columnwidth}{l c >{\centering\arraybackslash}X >{\raggedleft\arraybackslash}X >{\raggedleft\arraybackslash}X >{\raggedleft\arraybackslash}X >{\raggedleft\arraybackslash}X >{\raggedleft\arraybackslash}X}
\toprule
\textbf{Model} & \textbf{EX} &
\multicolumn{2}{c}{\textbf{E2E (s)}} &
\multicolumn{4}{c}{\textbf{\% of E2E Time}} \\
\cmidrule(lr){3-4} \cmidrule(lr){5-8}
& & \textbf{Mean} & \textbf{$\sigma$} & \textbf{list} & \textbf{schema} & \textbf{check} & \textbf{run} \\
\midrule
GPT-4o & 0.93 & 6.55 & 2.08 & 9.22 & 13.15 & 62.08 & 13.64 \\
\rowcolor{lightgray}Gemini 3 Flash & 1.00 & 8.37 & 2.61 & 9.74 & 10.15 & 66.76 & 11.71 \\
GPT-5.2 & 0.69 & 8.44 & 2.74 & 13.18 & 21.36 & 48.88 & 14.88 \\
\rowcolor{lightgray}Gemini 2.5 Flash & 0.95 & 9.18 & 3.21 & 11.70 & 10.53 & 66.07 & 10.11 \\
Claude Opus 4.5 & 1.00 & 11.40 & 2.82 & 16.90 & 18.57 & 42.32 & 20.93 \\
\rowcolor{lightgray}Claude Opus 4.6 & 1.00 & 12.60 & 2.09 & 17.18 & 18.11 & 42.70 & 20.93 \\
Kimi K2.5 & 0.98 & 13.61 & 5.72 & 9.99 & 14.35 & 62.89 & 11.74 \\
\rowcolor{lightgray}GPT-5 & 0.88 & 15.45 & 8.63 & 7.91 & 11.09 & 72.04 & 7.98 \\
Gemini 3 Pro & 1.00 & 54.55 & 45.34 & 18.93 & 18.85 & 45.30 & 16.65 \\
\rowcolor{lightgray}GLM-5 & 1.00 & 79.63 & 50.57 & 11.77 & 10.60 & 66.91 & 10.53 \\
\bottomrule
\end{tabularx}
\end{table}

For better observability, we breakdown execution time, aggregating Observation-Thought-Action iterations that call the same tool into the \textit{stage} abstraction. Common patterns emerge across models: the \texttt{check\_query} stage dominates end-to-end time in all LLMs, as expected since it runs within the LLM rather than the local Spark session. Yet percentages vary widely (e.g., a 23\% spread between GPT-5 and GPT-5.2).

These results suggest stage-specific optimization via model selection. For instance, the time split between \texttt{list\_tables} and \texttt{run\_query} differs by model. \textbf{Overall, smart per-stage model assignment (as in model ensembles~\cite{cheng2025barbariansgateaiupending}) offers clear optimization potential}.

\subsection{Big SQL Metrics Zoom In on Performance}
\label{sec:evaluation-metrics}

In the context of Big Data, text-to-SQL metrics fail to inform model selection. Instead, a Big SQL lens differentiates LLMs more clearly, providing sharper selection criteria. Table~\ref{tab:time-var-combined} shows normalized VES and
VES* with respect to the best scoring LLM setup: GPT-4o. As shown in the table, the \textbf{VES* metric better discriminates accurate models (809.09\% dispersion range versus 54.93\% in VES)}. The main reason is that VES considers only query execution time and result accuracy~\cite{hong2025database-interfaces}. However, modern LLMs excel at text-to-SQL: their generated queries often functionally resemble the gold query (even if not identical) yielding similar execution times. At high accuracy levels, this hinders model discrimination. Our proposed metric addresses this limitation by breaking ties, favoring LLMs that combine efficient agentic interaction with minimal projection overhead (i.e., avoiding superfluous columns).

\begin{table}[ht]
\centering
\caption{VES and VES* for the selected BIRD queries. Both metrics are normalized to the best scoring LLM. LLMs are displayed in descending order of VES*.}
\label{tab:time-var-combined}
\setlength{\tabcolsep}{2.5pt}
\footnotesize
\begin{tabularx}{\columnwidth}{l >{\centering\arraybackslash}X >{\centering\arraybackslash}X >{\raggedleft\arraybackslash}X >{\raggedleft\arraybackslash}X >{\raggedleft\arraybackslash}X >{\raggedleft\arraybackslash}X}
\toprule
\textbf{Model} & \textbf{VES (norm)} & \textbf{VES* (norm)} & \multicolumn{4}{c}{\textbf{Time Variation (x)}} \\
\cmidrule(lr){4-7}
& & & \textbf{list} & \textbf{schema} & \textbf{check} & \textbf{run} \\
\midrule
GPT-4o & \textbf{1.00} & \textbf{1.00} & 1.00x & 1.00x & 1.00x & 1.00x \\
\rowcolor{lightgray}Gemini 3 Flash & 1.06 & 0.81 & 1.35x & 0.99x & 1.37x & 1.10x \\
Gemini 2.5 Flash & 1.00 & 0.78 & 1.78x & 1.12x & 1.49x & 1.04x \\
\rowcolor{lightgray}Claude Opus 4.5 & 1.09 & 0.57 & 3.19x & 2.46x & 1.19x & 2.67x \\
Claude Opus 4.6 & 1.09 & 0.51 & 3.58x & 2.65x & 1.32x & 2.95x \\
\rowcolor{lightgray}GPT-5 & 0.89 & 0.46 & 2.02x & 1.99x & 2.74x & 1.38x \\
Kimi K2.5 & 1.05 & 0.45 & 2.25x & 2.27x & 2.10x & 1.79x \\
\rowcolor{lightgray}GPT-5.2 & 0.71 & 0.39 & 1.84x & 2.09x & 1.01x & 1.41x \\
Gemini 3 Pro & 1.06 & 0.23 & 17.10x & 11.95x & 6.08x & 10.17x \\
\rowcolor{lightgray}GLM-5 & 1.05 & 0.11 & 15.52x & 9.81x & 13.10x & 9.39x \\
\bottomrule
\end{tabularx}
\end{table}

\noindent\textbf{A high VES* effectively reflects both accurate and low latency models}, ranking GPT-4o highest (Table~\ref{tab:time-var-combined}).
It also reveals nuances, such as the Opus models benefiting from perfect accuracy despite incurring higher execution times. When normalizing execution times relative to the fastest model (GPT-4o), GPT-4o consistently ``wins'' across all stages, indicating a clear separation between``fast'' and ``slow'' model classes. VES is unable to capture this system behavior.






\begin{table}[ht]
\centering
\caption{VCES and CVQ for the selected BIRD queries. The VCES values are normalized to the best-performing LLM, and models are shown in descending order of VCES.}
\label{tab:cost-var-combined}
\setlength{\tabcolsep}{2.5pt}
\footnotesize
\begin{tabularx}{\columnwidth}{l >{\centering\arraybackslash}X >{\centering\arraybackslash}X >{\raggedleft\arraybackslash}X >{\raggedleft\arraybackslash}X >{\raggedleft\arraybackslash}X >{\raggedleft\arraybackslash}X}
\toprule
\textbf{Model} & \textbf{VCES norm. ($\$^{-1}$)} & \textbf{CVQ (\$)} & \multicolumn{4}{c}{\textbf{Cost Variation (x)}} \\
\cmidrule(lr){4-7}
& & & \textbf{list} & \textbf{schema} & \textbf{check} & \textbf{run} \\
\midrule
Gemini 3 Flash & \textbf{1.00} & 0.0044 & 1.00x & 1.00x & 1.00x & 1.00x \\
\rowcolor{lightgray}Gemini 2.5 Flash & 0.85 & 0.0053 & 1.36x & 1.18x & 1.12x & 0.98x \\
GPT-4o & 0.55 & 0.0107 & 2.14x & 2.93x & 2.10x & 2.64x \\
\rowcolor{lightgray}Kimi K2.5 & 0.53 & 0.0047 & 1.07x & 1.48x & 0.99x & 1.05x \\
GPT-5.2 & 0.25 & 0.0124 & 2.62x & 4.07x & 1.42x & 2.46x \\
\rowcolor{lightgray}GPT-5 & 0.24 & 0.0118 & 1.92x & 2.58x & 2.55x & 1.61x \\
Claude Opus 4.5 & 0.08 & 0.0388 & 15.30x & 16.13x & 5.59x & 15.76x \\
\rowcolor{lightgray}Claude Opus 4.6 & 0.08 & 0.0359 & 14.39x & 14.56x & 5.22x & 14.58x \\
Gemini 3 Pro & 0.06 & 0.0220 & 9.71x & 9.29x & 3.39x & 7.11x \\
\rowcolor{lightgray}GLM-5 & 0.05 & 0.0129 & 3.54x & 3.06x & 2.94x & 2.63x \\
\bottomrule
\end{tabularx}
\end{table}

\noindent \textbf{VCES matches the discriminative power of VES*} while incorporating cost, as it factors in per-token billing and the expenses of suboptimal query execution. Table~\ref{tab:cost-var-combined} presents VCES per LLM and identifies Gemini 3 Flash as the most cost-efficient option due to its low per-token pricing\footnote{0.5$/3.0$ per input/output token for Gemini 3 Flash~\cite{GoogleGeminiPricing2026}, vs. 2.5$/10.0$ for GPT-4o~\cite{OpenAIPricing2026}.}. Ultimately, \textbf{VCES complements VES* by enabling the selection of cost-efficient LLM setups, thus addressing a significant gap in text-to-SQL metrics that directly ignore cost}. We attach CVQ to better support our finding: GPT-4o more than doubles the per-query cost of Gemini 3 Flash due to its lower accuracy, which leads to more failed queries and higher token consumption.

As with execution time results, the leading cost performer in Table~\ref{tab:cost-var-combined} dominates across nearly all stages. Interestingly, the results reveal a clear trade-off between \textbf{latency-optimal} and\textbf{ cost-efficient} models: for example, GPT-4o ranks first in execution speed but achieves only half the cost-efficiency of Gemini 3 Flash, which is slightly slower but much cheaper. This distinction suggests that cost-efficient models may be preferable for stages with relaxed  latency requirements, while VES*-optimal models are ideal when latency is critical.

\subsection{The Aftermath of Data Scale}
\label{sec:evaluation-scale}

\begin{figure}[htbp]
    \centering
    \includegraphics[width=\linewidth]{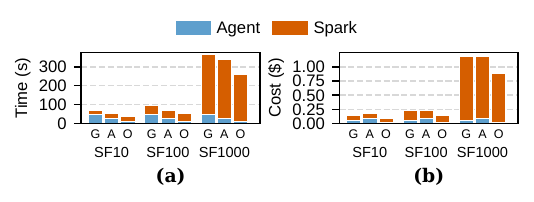}
    \caption{Breakdown of agent execution time (a) and cost (b) across different scale factors (SF) for TPC-H Query 21. Each bar series represents a specific model: Gemini 3 Pro (G), Claude Opus 4.5 (A), and GPT-5.2 (O).}
    \label{fig:tpc-h-21}
\end{figure}

Text-to-SQL metrics do not capture data scale. However, data scale plays a key role in text-to-Big SQL, as shown in~Figure~\ref{fig:tpc-h-21}.
The key observation is that both ends can hinder interactive analytics, as seen across all three models. At small scale factors (SFs), agent interactions dominate the overall execution time, while at large SFs, fast responses are limited by the long query execution times; the execution time of TPC-H Query 21 increases by 13.3$\times$ when scaling from SF 10 to SF 1000 on the same cluster~\footnote{Tests ran on a Amazon EMR~\cite{aws_emr_2026} cluster with \texttt{r5b.xlarge} master/core nodes, 32 core nodes (4 vCPUs each), and 128~GB gp3 EBS volumes.}. Even with highly optimized query engines, the end-to-end performance may remain constrained by the latency introduced through LLM–agent–tool interactions.

\begin{figure}[htbp]
    \centering
    \includegraphics[width=0.7\linewidth]{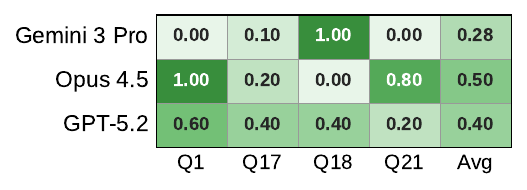}
    \caption{Execution accuracy across three models for TPC-H queries 1, 17, 18, and 21. Accuracy is computed as the average value of Equation~\ref{eq:accuracy-indicator} over 10 zero-shot translations of each corresponding natural language query.}
    \label{fig:tpc-h-ex}
\end{figure}

The remaining key question is the extent to which a new metric is needed to capture data scaling effects.
To examine this, we chose four TPC-H queries where LLMs deliver poor accuracy (Figure~\ref{fig:tpc-h-ex}): three complex ones (Q17, Q18, Q21) with nested subqueries, plus one simple single-table query (Q1). We used TPC-H business questions as NL inputs and official SQL as ground truth (see Appendix for details).


\begin{figure}[htbp]
\centering
\begin{subfigure}{0.48\columnwidth}
\centering
\includegraphics[width=\linewidth]{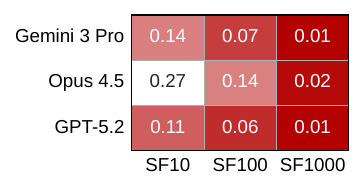}
\caption{VES}
\label{fig:tpc-h-ves}
\end{subfigure}
\hfill 
\begin{subfigure}{0.48\columnwidth}
\centering
\includegraphics[width=\linewidth]{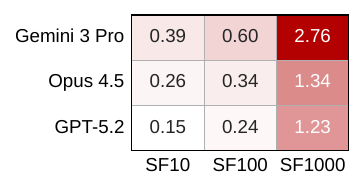}
\caption{CVQ}
\label{fig:tpc-h-svq}
\end{subfigure}
\caption{Text-to-SQL (a) and text-to-Big SQL (b) metrics across three models and scale factors, averaged for TPC-H queries 1, 17, 18, and 21.}
\label{fig:tpc-h-scale}
\end{figure}

As illustrated in Figure~\ref{fig:tpc-h-scale}, \textbf{VES (which incorporates SQL execution time) yields constant relationships between models across scale factors. In contrast, CVQ better captures each model's potential cost loss at varying scales: less accurate models (e.g., Gemini 3 Pro here) pose greater risk at large scale factors, where query errors incur substantially higher costs.} Consequently, failing a query at SF 1000 is far more expensive than at SF 10. Even a modest accuracy gap (such as the 10\% difference between Opus 4.5 and GPT-5.2) becomes critically amplified at higher scales.

Our evaluation clarifies the interpretation of text-to-Big SQL metrics. VES* and VCES provide practical assessments at fixed data scale factors, as they normalize generated SQL execution time and cost to the ground truth. In contrast, CVQ serves as an essential complement by quantifying the amplified impact of query inaccuracies as data scales.

\section{Discussion}
\label{sec:discussion}

We discuss our results in two dimensions. First, we elaborate on how our contributions should be interpreted and extended to provide more representative benchmarking for production contexts. Second, we identify promising research directions emerging from these initial steps toward Text-to-Big SQL.

\subsection{Benchmarking Real-World Deployments}

Text-to-SQL benchmarks offer a practical and reproducible means of evaluating translators, yet they fail to faithfully capture business and production contexts. Constructing a realistic Text-to-Big SQL benchmark requires accounting for the specific characteristics of production Big Data environments.

In benchmarks, we assume the golden query is the sole correct translation of its corresponding natural language query. However, in Big Data, a correct query can take several semantically equivalent forms, affecting physical plan quality, shuffled data size, scan efficiency, and therefore latency and resource consumption~\cite{song2026quitequeryrewriterules}.

The logical approach would designate the \textit{optimal} or \textit{worst-case} correct SQL as the golden query for each natural language query. Identifying these extremes is computationally infeasible due to the combinatorial explosion of possible query rewrite options, particularly for complex queries with multiple variables. Additionally, optimal queries are backend-dependent: a golden query for one system may be suboptimal for another based on pricing models, execution engines, or data partitioning strategies. Consequently, golden queries should serve not as universal baselines of optimality, but as reference points for relativizing translator efficiency.

Furthermore, text-to-SQL benchmarks execute each query an equal number of times to compute metrics~\cite{li2023bird,hong2025database-interfaces}. In practice, however, query frequencies follow skewed distributions where certain patterns dominate; for instance, certain operators may be significantly more frequent than others~\cite{vanRenen2023CAB}, wich complicates absolute performance assessments~\footnote{For instance, an abnormally frequent query structure accompanied by highly optimal golden queries would skew the overall benchmark assessment.}. Consequently, our method enables fair comparisons of text-to-SQL engines under identical conditions (same backend and golden query set) rather than providing absolute performance rankings.

Ideally, Text-to-Big SQL benchmarks should acknowledge this diversity of correct solutions by evaluating against multiple valid query variants rather than a single golden query. Meaningful comparison requires analyzing not merely syntactic structure, but the operator semantics of underlying physical execution plans, examined both pairwise and holistically across varying scale factors. We view this as a challenging challenge that merits broader future research.

\subsection{Future Opportunities in Text-to-Big SQL}

\begin{figure}[htbp]
    \centering
    \includegraphics[width=0.85\linewidth]{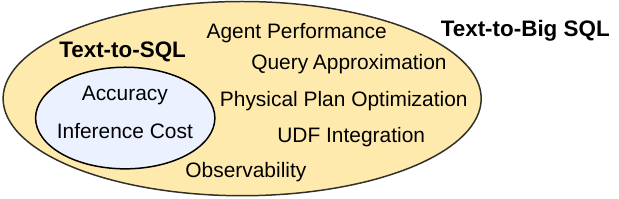}
    \caption{Challenges of Text-to-Big SQL include, but are not limited to, those of Text-to-SQL.}
     \label{fig:challenges}
\end{figure}

Based on our result, we identify several promising avenues in the broader text-to-Big SQL domain. Figure~\ref{fig:challenges} summarizes~our key insight: even if text-to-SQL reached 99\% accuracy since natural language ambiguity is unavoidable, text-to-Big SQL challenges would still remain.

\medskip

\noindent\textbf{Agent performance tuning.} Optimizing the internal stages of agents itself presents a significant research opportunity. Both our findings and prior work~\cite{cheng2025barbariansgateaiupending} demonstrate that both performance and cost could be improved by ``strategically'' assigning specialized models to different stages (e.g., navigating ``fast'' and` ``cheap'' models). However, existing physical plan optimization approaches for semantic operators incur latencies of tens of seconds~\cite{Russo_Liu_Sudhir_Vitagliano_Cafarella_Kraska_Madden_2026,zhu2025relationalsemanticawaremultimodalanalytics}, making them incompatible with interactive analytics. Adapting~these models to meet the latency constraints of interactive Big Data analytics remains a major open challenge.
\medskip

\noindent\textbf{Text-to-SQL must be optimized for large-scale query execution.} In Big Data, syntactically correct SQL may still be impractical if it triggers large shuffles, unnecessary joins, or full-table scans. Text-to-Big SQL must therefore optimize for both correctness and cost-efficient, large-scale execution, which remains an open challenge.

One promising direction is to leverage historical execution traces enriched with performance metrics and system-level quality indicators, such as VCES and CVQ.
By semantically matching newly generated
queries~\cite{fu2023catsql} to past executions \cite{wang2025andromeda}, the system can estimate expected cost and runtime before execution and proactively rewrite inefficient queries \cite{song2026quitequeryrewriterules,he2025aqorafastlearnedadaptive}. Incorporating physical plans and cost models~\cite{baldacci2019cost_model} further enables extrapolation across data scales, supporting scale-aware optimization rather than static SQL translation. Alternatively, the agent can be few-shot with similar past queries and their Big SQL metrics to infer optimized SQL.

Beyond exact execution, Big Data systems often rely on approximate queries to trade precision for performance~\cite{chaudhuri17}. However, current text-to-SQL models rarely reason about such semantic alternatives. A text-to-Big SQL framework should instead consider approximate joins, sampling-based aggregations, or sketch-based summaries when they satisfy user intent while significantly reducing execution cost. User-provided QoS annotations (e.g., ``run this fast'') could further guide optimization.
\medskip

\noindent\textbf{User-defined functions (UDFs).}  Another key challenge in text-to-Big SQL arises from UDFs. Big Data engines like Spark, Athena, and BigQuery often leverage custom UDFs, which are not fully expressible in standard SQL. As a result, text-to- Big SQL solutions
must produce UDF-compatible SQL or hybrid code, for example, combining SQL with Spark DataFrames, which goes beyond classical text-to-SQL.

\section{Related Work}

While existing benchmarks have advanced text-to-SQL research, they predominantly target moderate-scale relational databases and focus on translation accuracy in isolation. For instance, Spider 2.0~\cite{spider202025}, BIRD~\cite{li2024dawn}, and FinSQL~\cite{zhang2024finsql} emphasize complex upstream data sources but overlook the downstream cost implications of executing generated queries at scale. Cost-aware studies~\cite{deochake2025costawaretexttosqlempiricalstudy,zhang2024finsql} evaluate LLM performance while treating accuracy as binary. This gap is increasingly relevant as production tools like BigQuery's Generative AI~\cite{google2026bigquerygenai} already integrate text-to-SQL with Big Data workloads, making both translation quality and execution efficiency critical.

Beyond binary accuracy, recent works have explored fine-grained correctness metrics. Some of them~\cite{spider202025,Pinna2025redefining} move beyond all-or-nothing evaluation, recognizing that partial correctness matters in practice. Others apply similar reasoning to improve column linking in query generation~\cite{knapsack2025}. However, none of these efforts target the Big Data domain, where SQL execution directly influences cost and latency. We address this by integrating partial correctness with SQL performance metrics.

LLMs have proven effective at text-to-SQL translation~\cite{li2025OmniSQL,li2024dawn}, yet leveraging them in practice requires agent-based scaffolding to interact with user-specific databases~\cite{luo2025state-of-the-art}. While some frameworks integrate agents with Big Data systems~\cite{heng2025langchain-hpc}, their text-to-SQL effectiveness remains unexplored. Conversely, agent benchmarks such as BIRD-Interact~\cite{huo2025birdinteractreimaginingtexttosqlevaluation} evaluate interactivity and tool use but neglect query execution time and cost. This work introduces an evaluation framework that jointly assesses agent interactivity, translation accuracy, and downstream job performance.

\section{Conclusion}

In this work, we have started addressing the surprisingly underexplored domain of ``text-to-Big SQL'': the integration of text-to-SQL systems within Big Data engines.
We have shown that standard text-to-SQL benchmarks fall short and, for the first time, introduced text-to-Big SQL measures that reflect the true interaction between agents, LLMs, Big Data systems, and data scale. We have evaluated state-of-the-art production LLMs to demonstrate that our metrics provide an effective assessment framework for  text-to-Big SQL. Our work exposes new real-world challenges and guides future research in text-to-Big SQL.

\begin{acks}
This work has been partly funded by the EU Horizon programme, grant no. 101092646 (CloudSkin), and by the Spanish MICIU/AEI, grant no. PID2023-148202OB-C21. Germán T. Eizaguirre is recipient of a pre-doctoral FPU grant from the Spanish Ministry of Universities (ref. FPU21/00630).
\end{acks}

\bibliographystyle{ACM-Reference-Format}
\bibliography{sample-base}

\onecolumn
\appendix
\clearpage 
\section*{Appendix}

\section{Text-to-SQL formulas}

A text-to-SQL benchmark suite includes a set of triples containing a\textit{ natural language} (NL) query, a \textit{golden query} in SQL ($Q^n$), and a \textit{ground truth} ($V^n$) result~\cite{li2023bird}. During evaluation,  system performance is measured by comparing the generated SQL ($\hat{Q}^n$) to the golden query, and the resulting output ($\hat{V}^n$) to $V^n$~\cite{hong2025database-interfaces}.

Below we list standard formulas for text-to-SQL accuracy evaluation metrics, which serve as the foundation for the text-to-Big SQL metrics introduced in Section~\ref{sec:metrics}.

\subsection{Exact Matching (EM)}

\begin{equation}
EM = \frac{1}{N} \sum_{i=1}^{N} \mathbb{I} \left( Q_i, \hat{Q}_i \right)
\end{equation}

where

\begin{equation} \label{eq:em}
\mathbb{I}(Q_i, \hat{Q}_i) = \begin{cases}
1, & Q_i = \hat{Q}_i \\
0, & Q_i \neq \hat{Q}_i
\end{cases}
\end{equation}

The $Q_i = \hat{Q}_i$ operation compares two queries to determine their equivalence. Query equivalence depends on context. Therefore, multiple methods have been developed to address it. A common metric is the Spider exact matching accuracy~\cite{yu2019spiderlargescalehumanlabeleddataset}, which verifies whether generated SQL queries match gold-standard references in both structure and specific components. To achieve this, this metric decomposes a SQL query into its constituent clauses, such as \texttt{SELECT}, \texttt{WHERE}, \texttt{HAVING}, \texttt{GROUP BY}, and \texttt{ORDER BY}, and performs set-based comparisons. This approach effectively ignores ordering differences, such as varying sequences in conditions, to focus on \textbf{semantic equivalence}.

\subsection{Execution Accuracy (EA)}

\begin{equation}
EA = \frac{1}{N} \sum_{i=1}^{N} \mathbb{I} \left( V_i, \hat{V}_i \right)
\end{equation}

where

\begin{equation} \label{eq:ea-original}
\mathbb{I}(V_i, \hat{V}_i) = \begin{cases}
1, & V_i = \hat{V}_i \\
0, & V_i \neq \hat{V}_i
\end{cases}
\end{equation}

Usually, $V_i = \hat{V}_i$ uses a set-based equality: rows and columns must match exactly, ignoring order unless sorting is specified

\subsection{Valid Efficiency Score}

\begin{equation} \label{eq:ves}
\text{VES} = \frac{1}{N} \sum_{i=1}^{N} \left( \mathbb{I}(V_i, \hat{V}_i) \cdot \frac{T_{gold}}{T_{gen}} \right)
\end{equation}

Where $T_{gold}$ and $T_{gen}$ are the execution times of the golden query and the generated SQL, respectively.


\section{BIRD Evaluation}

\subsection{Metrics Breakdown for BIRD}

Our BIRD evaluation averages each metric over 50 iterations. Table~\ref{tab:appendix-bird-queries} provides a breakdown of the resulting Text-to-(Big) SQL metrics per query and model.

\begin{table*}[!tbp]
\caption{Per-query BIRD metrics by model (mean $\pm$ std across runs). VES, VES*, and VCES follow the paper definitions; CVQ reports the expected cost per valid query.}\label{tab:appendix-bird-queries}
\vspace{-0.3cm}
\footnotesize
\setlength{\tabcolsep}{1.5pt}
\renewcommand{\arraystretch}{0.55}
\centering
\begin{minipage}[t]{0.325\textwidth}
\centering
\begin{tabular}{@{}c l c c c c c@{}}
\toprule
\textbf{QID} & \textbf{Model} & \textbf{Acc.} & \textbf{VES} & \textbf{VES*} & \textbf{VCES} & \textbf{CVQ} \\
\midrule
\multirow{11}{*}{61} & GPT-4o & 1.00 & 1.08 & 0.011 & 1.11 & 0.010 \\
  & GPT-5.2 & 1.00 & 0.99 & 0.006 & 0.69 & 0.009 \\
  & Claude 4.6 & 1.00 & 1.06 & 0.006 & 0.20 & 0.030 \\
  & Gemini 3 Flash & 1.00 & 1.06 & 0.006 & 0.91 & 0.006 \\
  & Claude 4.5 & 1.00 & 1.07 & 0.006 & 0.14 & 0.040 \\
  & Kimi K2.5 & 1.00 & 1.08 & 0.005 & 1.11 & 0.005 \\
  & DeepSeek Chat & 0.86 & 0.90 & 0.003 & 1.73 & 0.002 \\
  & Gemini 2.5 Flash & 0.60 & 0.62 & 0.003 & 0.40 & 0.013 \\
  & GPT-5 & 1.00 & 1.06 & 0.003 & 0.23 & 0.013 \\
  & Gemini 3 Pro & 1.00 & 1.10 & 0.002 & 0.07 & 0.024 \\
  & GLM-5 & 1.00 & 1.02 & 0.001 & 0.05 & 0.015 \\
\midrule
\multirow{11}{*}{606} & Gemini 3 Flash & 1.00 & 0.98 & 0.012 & 1.87 & 0.006 \\
  & Claude 4.5 & 1.00 & 1.01 & 0.011 & 0.30 & 0.038 \\
  & Gemini 2.5 Flash & 1.00 & 0.97 & 0.011 & 1.62 & 0.007 \\
  & GPT-4o & 0.52 & 0.51 & 0.011 & 1.10 & 0.019 \\
  & Claude 4.6 & 1.00 & 0.98 & 0.010 & 0.30 & 0.035 \\
  & Kimi K2.5 & 1.00 & 1.00 & 0.010 & 2.14 & 0.005 \\
  & DeepSeek Chat & 0.92 & 0.83 & 0.006 & 2.87 & 0.002 \\
  & Gemini 3 Pro & 1.00 & 0.98 & 0.003 & 0.14 & 0.022 \\
  & GLM-5 & 1.00 & 0.96 & 0.001 & 0.06 & 0.022 \\
  & GPT-5.2 & 0.04 & 0.02 & 0.001 & 0.08 & 0.198 \\
  & GPT-5 & 0.06 & 0.03 & 0.000 & 0.02 & 0.253 \\
\midrule
\multirow{11}{*}{645} & GPT-4o & 1.00 & 0.98 & 0.012 & 1.31 & 0.009 \\
  & GPT-5.2 & 1.00 & 0.99 & 0.011 & 1.95 & 0.006 \\
  & Gemini 3 Flash & 1.00 & 0.99 & 0.010 & 2.86 & 0.003 \\
  & Gemini 2.5 Flash & 1.00 & 0.99 & 0.009 & 2.40 & 0.004 \\
  & Kimi K2.5 & 1.00 & 1.00 & 0.006 & 1.57 & 0.004 \\
  & Claude 4.5 & 1.00 & 0.99 & 0.006 & 0.15 & 0.037 \\
  & GPT-5 & 1.00 & 0.99 & 0.006 & 0.73 & 0.008 \\
  & Claude 4.6 & 1.00 & 1.02 & 0.006 & 0.16 & 0.034 \\
  & DeepSeek Chat & 1.00 & 1.00 & 0.004 & 1.96 & 0.002 \\
  & Gemini 3 Pro & 1.00 & 0.96 & 0.002 & 0.09 & 0.020 \\
  & GLM-5 & 1.00 & 0.98 & 0.001 & 0.15 & 0.009 \\
\bottomrule
\end{tabular}
\end{minipage}%
\hfill
\begin{minipage}[t]{0.325\textwidth}
\centering
\begin{tabular}{@{}c l c c c c c@{}}
\toprule
\textbf{QID} & \textbf{Model} & \textbf{Acc.} & \textbf{VES} & \textbf{VES*} & \textbf{VCES} & \textbf{CVQ} \\
\midrule
\multirow{11}{*}{776} & GPT-4o & 0.98 & 0.96 & 0.009 & 0.86 & 0.011 \\
  & Gemini 3 Flash & 1.00 & 0.99 & 0.009 & 2.19 & 0.004 \\
  & Gemini 2.5 Flash & 1.00 & 1.00 & 0.008 & 1.94 & 0.004 \\
  & GPT-5.2 & 1.00 & 1.09 & 0.006 & 0.63 & 0.010 \\
  & Claude 4.5 & 1.00 & 1.08 & 0.006 & 0.15 & 0.040 \\
  & Claude 4.6 & 1.00 & 1.13 & 0.005 & 0.14 & 0.037 \\
  & Kimi K2.5 & 0.96 & 0.94 & 0.005 & 0.94 & 0.005 \\
  & GPT-5 & 1.00 & 1.04 & 0.004 & 0.49 & 0.009 \\
  & DeepSeek Chat & 0.50 & 0.49 & 0.002 & 0.77 & 0.004 \\
  & Gemini 3 Pro & 1.00 & 1.00 & 0.001 & 0.06 & 0.024 \\
  & GLM-5 & 1.00 & 1.12 & 0.001 & 0.11 & 0.011 \\
\midrule
\multirow{11}{*}{607} & GPT-4o & 1.00 & 0.99 & 0.011 & 1.36 & 0.008 \\
  & Gemini 3 Flash & 1.00 & 0.98 & 0.009 & 2.89 & 0.003 \\
  & GPT-5.2 & 1.00 & 0.96 & 0.009 & 1.58 & 0.006 \\
  & Gemini 2.5 Flash & 1.00 & 0.96 & 0.009 & 2.47 & 0.003 \\
  & Claude 4.5 & 1.00 & 1.00 & 0.006 & 0.16 & 0.036 \\
  & GPT-5 & 1.00 & 1.01 & 0.005 & 0.72 & 0.007 \\
  & Kimi K2.5 & 1.00 & 0.99 & 0.005 & 1.27 & 0.004 \\
  & Claude 4.6 & 1.00 & 0.98 & 0.005 & 0.14 & 0.034 \\
  & DeepSeek Chat & 1.00 & 0.98 & 0.004 & 2.08 & 0.002 \\
  & Gemini 3 Pro & 1.00 & 0.98 & 0.002 & 0.10 & 0.017 \\
  & GLM-5 & 1.00 & 0.99 & 0.002 & 0.17 & 0.009 \\
\midrule
\multirow{11}{*}{785} & Gemini 3 Flash & 0.98 & 0.98 & 0.007 & 1.97 & 0.004 \\
  & GPT-4o & 0.90 & 0.88 & 0.007 & 0.66 & 0.012 \\
  & GPT-5.2 & 1.00 & 0.59 & 0.007 & 0.89 & 0.008 \\
  & Gemini 2.5 Flash & 1.00 & 0.96 & 0.007 & 1.55 & 0.004 \\
  & Claude 4.5 & 1.00 & 0.99 & 0.005 & 0.12 & 0.039 \\
  & Kimi K2.5 & 1.00 & 0.97 & 0.004 & 0.84 & 0.005 \\
  & Claude 4.6 & 1.00 & 0.99 & 0.004 & 0.11 & 0.037 \\
  & GPT-5 & 0.96 & 0.53 & 0.004 & 0.41 & 0.010 \\
  & Gemini 3 Pro & 1.00 & 0.97 & 0.001 & 0.06 & 0.022 \\
  & GLM-5 & 1.00 & 0.74 & 0.001 & 0.11 & 0.010 \\
  & DeepSeek Chat & 0.26 & 0.16 & 0.001 & 0.39 & 0.006 \\
\bottomrule
\end{tabular}
\end{minipage}%
\hfill
\begin{minipage}[t]{0.325\textwidth}
\centering
\begin{tabular}{@{}c l c c c c c@{}}
\toprule
\textbf{QID} & \textbf{Model} & \textbf{Acc.} & \textbf{VES} & \textbf{VES*} & \textbf{VCES} & \textbf{CVQ} \\
\midrule
\multirow{11}{*}{813} & GPT-4o & 0.98 & 1.46 & 0.016 & 1.57 & 0.011 \\
  & Gemini 3 Flash & 1.00 & 1.47 & 0.015 & 3.21 & 0.005 \\
  & GPT-5.2 & 1.00 & 1.41 & 0.015 & 1.90 & 0.008 \\
  & Gemini 2.5 Flash & 0.96 & 1.34 & 0.012 & 2.21 & 0.006 \\
  & Claude 4.5 & 1.00 & 1.51 & 0.011 & 0.28 & 0.040 \\
  & Claude 4.6 & 1.00 & 1.41 & 0.009 & 0.22 & 0.038 \\
  & Kimi K2.5 & 0.94 & 1.36 & 0.008 & 1.58 & 0.006 \\
  & GPT-5 & 1.00 & 1.43 & 0.008 & 0.79 & 0.010 \\
  & Gemini 3 Pro & 1.00 & 1.45 & 0.004 & 0.15 & 0.024 \\
  & GLM-5 & 1.00 & 1.52 & 0.002 & 0.11 & 0.015 \\
  & DeepSeek Chat & 0.18 & 0.26 & 0.001 & 0.89 & 0.007 \\
\midrule
\multirow{11}{*}{895} & GPT-4o & 1.00 & 0.95 & 0.073 & 5.61 & 0.013 \\
  & Gemini 2.5 Flash & 1.00 & 0.98 & 0.058 & 10.15 & 0.006 \\
  & Gemini 3 Flash & 1.00 & 0.96 & 0.054 & 10.71 & 0.005 \\
  & GPT-5 & 0.94 & 0.90 & 0.037 & 2.70 & 0.015 \\
  & Claude 4.5 & 1.00 & 0.95 & 0.033 & 0.61 & 0.053 \\
  & Claude 4.6 & 1.00 & 0.96 & 0.031 & 0.65 & 0.048 \\
  & Kimi K2.5 & 1.00 & 0.97 & 0.024 & 3.82 & 0.006 \\
  & Gemini 3 Pro & 1.00 & 0.97 & 0.020 & 0.77 & 0.026 \\
  & GLM-5 & 1.00 & 0.98 & 0.007 & 0.52 & 0.014 \\
  & DeepSeek Chat & 0.28 & 0.26 & 0.006 & 3.24 & 0.007 \\
  & GPT-5.2 & 0.12 & 0.11 & 0.005 & 0.29 & 0.139 \\
\midrule
\multirow{11}{*}{968} & GPT-4o & 1.00 & 0.92 & 0.003 & 0.39 & 0.009 \\
  & Claude 4.5 & 1.00 & 0.93 & 0.003 & 0.12 & 0.025 \\
  & Gemini 2.5 Flash & 1.00 & 0.92 & 0.003 & 0.82 & 0.004 \\
  & Gemini 3 Flash & 1.00 & 0.87 & 0.003 & 0.81 & 0.003 \\
  & Claude 4.6 & 1.00 & 0.95 & 0.002 & 0.07 & 0.031 \\
  & Kimi K2.5 & 0.96 & 0.85 & 0.002 & 0.54 & 0.004 \\
  & GPT-5 & 0.96 & 0.81 & 0.002 & 0.21 & 0.008 \\
  & DeepSeek Chat & 0.74 & 0.65 & 0.001 & 0.47 & 0.002 \\
  & Gemini 3 Pro & 1.00 & 0.87 & 0.001 & 0.04 & 0.018 \\
  & GLM-5 & 0.98 & 0.90 & 0.000 & 0.02 & 0.011 \\
  & GPT-5.2 & 0.04 & 0.04 & 0.000 & 0.02 & 0.151 \\
\bottomrule
\end{tabular}
\end{minipage}
\end{table*}

\subsection{Translation Error Analysis}

In the BIRD evaluation we executed 10 queries\footnote{Query 886 was excluded from the main paper results since the Equation~\ref{eq:ea-original} results are low for most models. The query asks \texttt{"Which year has the most number of races? The most number of races refers to max(round)"}; most models returned the superfluous column \texttt{max(round)} instead of only the year.} on 11 models\footnote{DeepSeek Chat is omitted in the main paper results due to its comparatively poor performance.} with 50 zero-shot iterations of the translation for each, resulting in 5,500 total query executions.  Out of these, 930 translations produced a result different from the expected output, yielding an Equation~\ref{eq:ea-original} result of 0. Table~\ref{tab:failed-iterations-per-model} presents the per-model aggregate distribution of the errors.

\begin{figure}[ht]
\centering
\begin{lstlisting}[language=SQL,basicstyle=\ttfamily\footnotesize\color{SeaGreen},breaklines=true,breakatwhitespace=true,columns=fullflexible,caption={Correct query for Q886},label={lst:q886-correct}]
SELECT year
FROM races
GROUP BY year
ORDER BY COUNT(round) DESC
LIMIT 1
\end{lstlisting}

\begin{lstlisting}[language=SQL,basicstyle=\ttfamily\footnotesize\color{BrickRed},breaklines=true,breakatwhitespace=true,columns=fullflexible,caption={Incorrect query for Q886 (translated by Opus 4.5)},label={lst:q886-incorrect}]
SELECT year, MAX(round) as max_races
FROM races
GROUP BY year
ORDER BY max_races DESC
LIMIT 1
\end{lstlisting}
\vspace{-12pt}
\caption{Correct and incorrect SQL examples for query 886.}
\label{fig:q886-correct-vs-incorrect}
\end{figure}

\begin{table}[htbp]
\caption{Total number of incorrect translations (Eq.\ref{eq:accuracy-indicator}$==0$) per model. Each model was evaluated on all 10 BIRD queries with 50 translations per query (500 iterations per model).}
\label{tab:failed-iterations-per-model}
\small
\begin{tabular}{l c}
\toprule
\textbf{Model} & \textbf{Incorrect Translations} \\
\midrule
DeepSeek Chat & 214 \\
GPT-5.2 & 190 \\
GPT-5 & 104 \\
GPT-4o & 78 \\
Gemini 2.5 Flash & 70 \\
Kimi K2.5 & 57 \\
Gemini 3 Flash & 51 \\
GLM-5 & 51 \\
Claude Opus 4.5 & 50 \\
Claude Opus 4.6 & 50 \\
Gemini 3 Pro & 15 \\
\bottomrule
\end{tabular}
\end{table}

\begin{figure}[!htbp]
    \centering
    \includegraphics[width=0.7\linewidth]{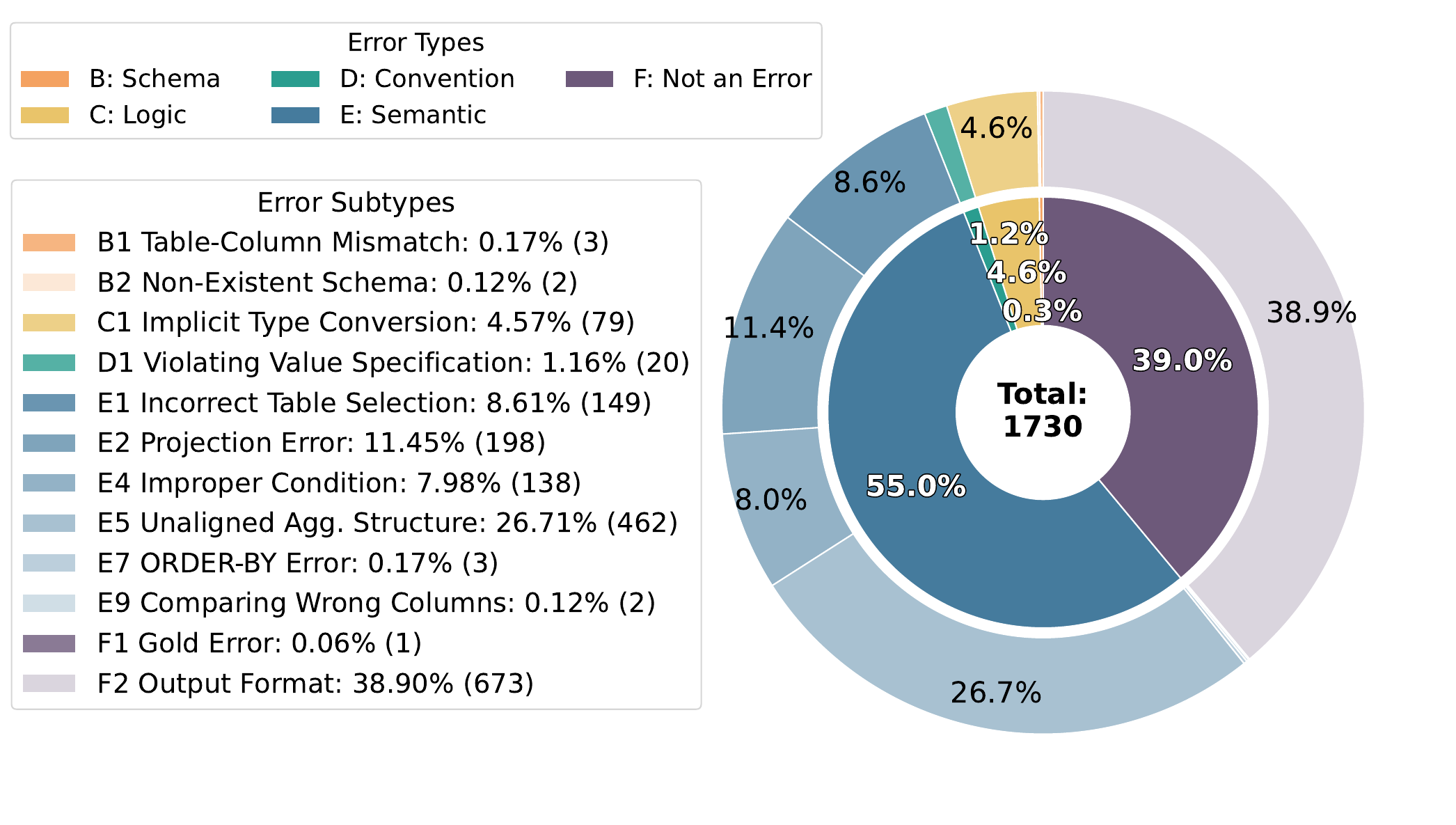}
    \caption{Distribution of identified text-to-SQL translation errors across all 930 incorrect BIRD query translations, categorized by the taxonomy proposed in~\cite{shen2025studyincontextlearningbasedtexttosqlerrors}}
    \label{fig:pie-chart}
\end{figure}

To provide a better analysis, we classify the obtained errors based on the taxonomy from~\cite{shen2025studyincontextlearningbasedtexttosqlerrors}. We use their proposed categories because they also focus on LLMs for text-to-SQL translation and provide an extensive categorization of possible outcomes. Each query with a result different from the expected output (i.e. `\textit{incorrect}') could be affected by one or more errors. We manually gathered 1,730 errors across the 930 incorrect queries. We depict the error distribution in Figure~\ref{fig:pie-chart} and a visual breakdown of the errors identified in Figure~\ref{fig:error-taxonomy}.

We extracted several interesting insights from the analysis. Within the \textbf{Output Format (F2)} errors, 627 (93.16\%) were due to additional columns, meaning the result contained the expected data accompanied by unnecessary columns. Of those, 77.51\% were due to just one additional column. Such results would be considered correct under our text-to-Big SQL metrics (Eq.~\ref{eq:accuracy-indicator}). For instance, query 886 is considered incorrect by classical text-to-SQL metrics primarily because of an additional column (see Listing~\ref{lst:q886-incorrect}). We consider our metric to better represent valid queries, as a user could easily differentiate the required column in a practical case. We also identified an LLM difficulty in differentiating MAX and COUNT cases. Among \textbf{Unaligned Aggregation Structure (E5)} errors, 15.80\% resulted from adding unnecessary \texttt{MAX} aggregations, while 5.41\% were due to unnecessary \texttt{COUNT}s; notably, 44.16\% involved mixing \texttt{MAX} and \texttt{COUNT} aggregators. Finally, in four queries, the agents ran an inspection operation (\texttt{DESCRIBE table} or \texttt{SHOW TABLES}) as the final query, a curious behavior despite the low number of cases.

We only tested the agent extensively on a subset of the BIRD benchmark; therefore, these results should not be considered absolute. However, we believe these insights provide valuable information for future research.

\section{TPC-H evaluation}

\subsection{Model and Query Selection}

We selected three last generation frontier models available at the time of our original experiments in January 2026: Gemini 3 Pro, Claude Opus 4.5, and GPT-5.2. We excluded additional models due to budget constraints, as a single TPC-H query run at the specified scale factor costs approximately \$1 within our proposed EMR cluster. Given 50 replicas per model and the orginal set of 10 models from the BIRD evaluation, the total cost for four TPC-H queries would reach roughly \$2,000, excluding lower scale factor tests. Because an exhaustive assessment of all available models and queries would not further clarify the primary contributions of this paper, we limited this evaluation to a representative subset.

Regarding query selection, we prioritized TPC-H queries that Claude Opus 4.5 could not translate correctly. We then verified that (1) accuracy remained imperfect in the other two models and (2) no model had memorized the TPC-H specification during training. While methods exist to prove memorization in production Large Language Models (LLMs), verifying the absolute absence of memorization remains an open challenge. As a workaround, we adapted the methodology from~\cite{ahmed2026extractingbooksproductionlanguage}~\footnote{We make the memorization test code available at \url{https://github.com/GEizaguirre/memorization-LLM-prod}}, and attempted to detect TPC-H memorization in the models; however, these efforts were unsuccessful.

\begin{figure*}[!t] 
\centering
\resizebox{0.6\textwidth}{!}{
\begin{tikzpicture}[
    x=1cm,
    y=1cm,
    font=\large,
    every node/.style={align=center},
]

\tikzset{
    root/.style={rectangle, rounded corners=2pt, draw=black, fill=black!80, text=white, font=\bfseries\small, minimum width=4.5cm, minimum height=0.75cm},
    catB/.style={rectangle, rounded corners=2pt, draw=black, fill=orange!30, font=\bfseries\scriptsize, minimum width=2.5cm, minimum height=0.75cm},
    catC/.style={rectangle, rounded corners=2pt, draw=black, fill=yellow!40, font=\bfseries\scriptsize, minimum width=2.5cm, minimum height=0.75cm},
    catD/.style={rectangle, rounded corners=2pt, draw=black, fill=green!30, font=\bfseries\scriptsize, minimum width=2.5cm, minimum height=0.75cm},
    catE/.style={rectangle, rounded corners=2pt, draw=black, fill=blue!30, font=\bfseries\scriptsize, minimum width=2.5cm, minimum height=0.75cm},
    catF/.style={rectangle, rounded corners=2pt, draw=black, fill=purple!30, font=\bfseries\scriptsize, minimum width=2.5cm, minimum height=0.75cm},
    errB/.style={rectangle, rounded corners=1.5pt, draw=black, fill=orange!10, text=black, text width=2.0cm, minimum height=0.65cm, inner sep=1.5pt, font=\tiny},
    errC/.style={rectangle, rounded corners=1.5pt, draw=black, fill=yellow!20, text=black, text width=2.0cm, minimum height=0.65cm, inner sep=1.5pt, font=\tiny},
    errD/.style={rectangle, rounded corners=1.5pt, draw=black, fill=green!10, text=black, text width=2.0cm, minimum height=0.65cm, inner sep=1.5pt, font=\tiny},
    errE/.style={rectangle, rounded corners=1.5pt, draw=black, fill=blue!10, text=black, text width=2.0cm, minimum height=0.65cm, inner sep=1.5pt, font=\tiny},
    errF/.style={rectangle, rounded corners=1.5pt, draw=black, fill=purple!10, text=black, text width=2.0cm, minimum height=0.65cm, inner sep=1.5pt, font=\tiny},
    edge/.style={thick, ->, >=stealth, draw=black!70},
}

\node[root] (root) at (0,0) {Error Taxonomy (1730 errors)};

\node[catB] (catB) at (-6.00,-1.50) {Schema\\(5)};
\node[catC] (catC) at (-3.00,-1.50) {Logic\\(79)};
\node[catD] (catD) at (0.00,-1.50) {Convention\\(20)};
\node[catE] (catE) at (3.00,-1.50) {Semantic\\(952)};
\node[catF] (catF) at (6.00,-1.50) {Not an Error\\(674)};

\draw[edge] (root.south) -- ++(0,-0.6) -| (catB.north);
\draw[edge] (root.south) -- ++(0,-0.6) -| (catC.north);
\draw[edge] (root.south) -- ++(0,-0.6) -| (catD.north);
\draw[edge] (root.south) -- ++(0,-0.6) -| (catE.north);
\draw[edge] (root.south) -- ++(0,-0.6) -| (catF.north);

\node[errB] (errB1) at (-6.00,-3.00) {\texttt{B1}\\Table-Column\\Mismatch\\(3)};
\node[errB] (errB2) at (-6.00,-4.00) {\texttt{B2}\\Non-Existent Schema\\(2)};
\node[errC] (errC1) at (-3.00,-3.00) {\texttt{C1}\\Implicit Type\\Conversion\\(79)};
\node[errD] (errD1) at (0.00,-3.00) {\texttt{D1}\\Violating Value\\Specification\\(20)};
\node[errE] (errE1) at (3.00,-3.00) {\texttt{E1}\\Incorrect Table\\Selection\\(149)};
\node[errE] (errE2) at (3.00,-4.00) {\texttt{E2}\\Projection Error\\(198)};
\node[errE] (errE4) at (3.00,-5.00) {\texttt{E4}\\Improper Condition\\(138)};
\node[errE] (errE5) at (3.00,-6.20) {\texttt{E5}\\Unaligned\\Aggregation\\Structure\\(462)};
\node[errE] (errE7) at (3.00,-7.40) {\texttt{E7}\\ORDER-BY Error\\(3)};
\node[errE] (errE9) at (3.00,-8.40) {\texttt{E9}\\Comparing Wrong\\Columns\\(2)};
\node[errF] (errF1) at (6.00,-3.00) {\texttt{F1}\\Gold Error\\(1)};
\node[errF] (errF2) at (6.00,-4.00) {\texttt{F2}\\Output Format\\(673)};

\draw[thick, draw=black!70] (catB.south) -- ++(0,-0.30) -| (-7.50,-3.00);
\draw[thick, draw=black!70] (-7.50,-3.00) -- (-7.50,-4.00);
\draw[edge] (-7.50,-3.00) -- (errB1.west);
\draw[edge] (-7.50,-4.00) -- (errB2.west);
\draw[edge] (catC.south) -- ++(0,-0.30) -| (errC1.north);
\draw[edge] (catD.south) -- ++(0,-0.30) -| (errD1.north);
\draw[thick, draw=black!70]  (catE.south) -- ++(0,-0.30) -| (4.50,-3.00);
\draw[thick, draw=black!70] (4.50,-3.00) -- (4.50,-8.40);
\draw[edge] (4.50,-3.00) -- (errE1.east);
\draw[edge] (4.50,-4.00) -- (errE2.east);
\draw[edge] (4.50,-5.00) -- (errE4.east);
\draw[edge] (4.50,-6.20) -- (errE5.east);
\draw[edge] (4.50,-7.40) -- (errE7.east);
\draw[edge] (4.50,-8.40) -- (errE9.east);
\draw[thick, draw=black!70]  (catF.south) -- ++(0,-0.30) -| (7.50,-3.00);
\draw[thick, draw=black!70]  (7.50,-3.00) -- (7.50,-4.00);
\draw[edge] (7.50,-3.00) -- (errF1.east);
\draw[edge] (7.50,-4.00) -- (errF2.east);

\end{tikzpicture}%
}
\caption{Taxonomy~\cite{shen2025studyincontextlearningbasedtexttosqlerrors} of SQL translation errors from our LLM agent.
Numbers in parentheses denote error counts.
Categories: A=Syntax, B=Schema, C=Logic, D=Convention, E=Semantic, F=Not an Error.
Total: 1730 errors across 5 categories.}
\label{fig:error-taxonomy}
\end{figure*}

\subsection{Metrics Breakdown for TPC-H}

Our TPC-H evaluation averages each metric over 50 iterations at SF 1. Table~\ref{tab:appendix-tpch-sf1-queries} provides a breakdown of the resulting Text-to-(Big) SQL metrics for each query and model.

\begin{table}[H]
\caption{Per-query metrics in TPC-H (SF1) by model (mean $\pm$ std across runs). VES, VES*, and VCES follow the paper definitions; CVQ reports the expected cost per valid query.}\label{tab:appendix-tpch-sf1-queries}
\small
\setlength{\tabcolsep}{1.5pt}
\begin{tabular}{@{}c l c c c c c@{}}
\toprule
\textbf{QID} & \textbf{Model} & \textbf{Acc.} & \textbf{VES} & \textbf{VES*} & \textbf{VCES} & \textbf{CVQ} \\
\midrule
\multirow{3}{*}{1} & GPT 5.2 & 0.60 $\pm$ 0.49 & 0.6507 $\pm$ 0.5318 & 0.2608 $\pm$ 0.2134 & 23.1794 $\pm$ 19.0188 & 0.0188 $\pm$ 0.0000 \\
 & Opus 4.5 & 1.00 $\pm$ 0.00 & 0.9567 $\pm$ 0.1057 & 0.2513 $\pm$ 0.0138 & 3.6030 $\pm$ 0.2671 & 0.0697 $\pm$ 0.0015 \\
 & Gemini 3 Pro (T) & 0.00 $\pm$ 0.00 & 0.0000 $\pm$ 0.0000 & 0.0000 $\pm$ 0.0000 & 0.0000 $\pm$ 0.0000 & -- \\
\midrule
\multirow{3}{*}{17} & Opus 4.5 & 0.20 $\pm$ 0.40 & 0.2028 $\pm$ 0.4235 & 0.0323 $\pm$ 0.0647 & 0.3865 $\pm$ 0.6072 & 0.4182 $\pm$ 0.0025 \\
 & Gemini 3 Pro (T) & 0.10 $\pm$ 0.30 & 0.1294 $\pm$ 0.3883 & 0.0140 $\pm$ 0.0420 & 0.3525 $\pm$ 0.7373 & 0.3973 $\pm$ 0.0000 \\
 & GPT-5.2 & 0.40 $\pm$ 0.49 & 0.0000 $\pm$ 0.0000 & 0.0000 $\pm$ 0.0000 & 0.0000 $\pm$ 0.0000 & 0.0168 $\pm$ 0.0004 \\
\midrule
\multirow{3}{*}{18} & Gemini 3 Pro (T) & 1.00 $\pm$ 0.00 & 1.3223 $\pm$ 0.1182 & 0.2449 $\pm$ 0.0232 & 6.4131 $\pm$ 0.8326 & 0.0382 $\pm$ 0.0019 \\
 & GPT-5.2 & 0.40 $\pm$ 0.49 & 0.2206 $\pm$ 0.6619 & 0.0711 $\pm$ 0.2133 & 7.5480 $\pm$ 17.8560 & 0.0236 $\pm$ 0.0030 \\
 & Opus 4.5 & 0.00 $\pm$ 0.00 & 0.0000 $\pm$ 0.0000 & 0.0000 $\pm$ 0.0000 & 0.0000 $\pm$ 0.0000 & -- \\
\midrule
\multirow{3}{*}{21} & Opus 4.5 & 0.80 $\pm$ 0.40 & 0.7732 $\pm$ 0.3889 & 0.2639 $\pm$ 0.1321 & 2.8673 $\pm$ 1.7429 & 0.1150 $\pm$ 0.0003 \\
 & GPT-5.2 & 0.20 $\pm$ 0.40 & 0.0000 $\pm$ 0.0000 & 0.0000 $\pm$ 0.0000 & 0.0000 $\pm$ 0.0000 & 0.0915 $\pm$ 0.0000 \\
 & Gemini 3 Pro (T) & 0.00 $\pm$ 0.00 & 0.0000 $\pm$ 0.0000 & 0.0000 $\pm$ 0.0000 & 0.0000 $\pm$ 0.0000 & -- \\
\bottomrule
\end{tabular}
\end{table}

We then execute the generated SQL queries across different scale factors and average the proposed metrics for all queries. These results are shown in Table~\ref{tab:appendix-tpch-scale-factors-combined}.

\begin{table}[H]
\caption{TPC-H metrics by query, model, and scale factor (mean $\pm$ std across runs). VES, VES*, and VCES follow the paper definitions; CVQ reports the expected cost per valid query.}
\label{tab:appendix-tpch-scale-factors-combined}
\small
\setlength{\tabcolsep}{2pt}
\begin{tabular}{@{}c c l c c c c@{}}
\toprule
\textbf{Query} & \textbf{SF} & \textbf{Model} & \textbf{VES} & \textbf{VES*} & \textbf{VCES} & \textbf{CVQ} \\
\midrule
\multirow{9}{*}{Q1}
  & \multirow{3}{*}{10} & Gemini 3 Pro & 0.0000 $\pm$ 0.0000 & 0.0000 $\pm$ 0.0000 & 0.0000 $\pm$ 0.0000 & -- \\
  &   & Opus 4.5 & 0.5668 $\pm$ 0.0000 & 0.2133 $\pm$ 0.0064 & 1.8351 $\pm$ 0.0783 & 0.1163 $\pm$ 0.0015 \\
  &   & GPT-5.2 & 0.3235 $\pm$ 0.2641 & 0.1857 $\pm$ 0.1517 & 3.0724 $\pm$ 2.5113 & 0.1007 $\pm$ 0.0000 \\
\cmidrule(l){2-7}
  & \multirow{3}{*}{100} & Gemini 3 Pro & 0.0000 $\pm$ 0.0000 & 0.0000 $\pm$ 0.0000 & 0.0000 $\pm$ 0.0000 & -- \\
  &   & Opus 4.5 & 0.2654 $\pm$ 0.0000 & 0.1494 $\pm$ 0.0031 & 0.8809 $\pm$ 0.0259 & 0.1697 $\pm$ 0.0015 \\
  &   & GPT-5.2 & 0.1599 $\pm$ 0.1306 & 0.1170 $\pm$ 0.0955 & 1.0526 $\pm$ 0.8598 & 0.1852 $\pm$ 0.0000 \\
\cmidrule(l){2-7}
  & \multirow{3}{*}{1000} & Gemini 3 Pro & 0.0000 $\pm$ 0.0000 & 0.0000 $\pm$ 0.0000 & 0.0000 $\pm$ 0.0000 & -- \\
  &   & Opus 4.5 & 0.0484 $\pm$ 0.0000 & 0.0424 $\pm$ 0.0002 & 0.0683 $\pm$ 0.0006 & 0.6199 $\pm$ 0.0015 \\
  &   & GPT-5.2 & 0.0289 $\pm$ 0.0236 & 0.0271 $\pm$ 0.0221 & 0.0478 $\pm$ 0.0390 & 0.9439 $\pm$ 0.0000 \\
\midrule
\multirow{9}{*}{Q17}
  & \multirow{3}{*}{10} & Gemini 3 Pro & 0.0519 $\pm$ 0.1556 & 0.0121 $\pm$ 0.0362 & 0.1148 $\pm$ 0.3444 & 0.8777 $\pm$ 0.0000 \\
  &   & Opus 4.5 & 0.1040 $\pm$ 0.2079 & 0.0284 $\pm$ 0.0570 & 0.1814 $\pm$ 0.3637 & 0.6704 $\pm$ 0.0026 \\
  &   & GPT-5.2 & 0.0000 $\pm$ 0.0000 & 0.0000 $\pm$ 0.0000 & 0.0000 $\pm$ 0.0000 & 0.1025 $\pm$ 0.0004 \\
\cmidrule(l){2-7}
  & \multirow{3}{*}{100} & Gemini 3 Pro & 0.0332 $\pm$ 0.0997 & 0.0107 $\pm$ 0.0320 & 0.0807 $\pm$ 0.2422 & 1.1488 $\pm$ 0.0000 \\
  &   & Opus 4.5 & 0.0669 $\pm$ 0.1337 & 0.0247 $\pm$ 0.0494 & 0.1335 $\pm$ 0.2674 & 0.8111 $\pm$ 0.0026 \\
  &   & GPT-5.2 & 0.0000 $\pm$ 0.0000 & 0.0000 $\pm$ 0.0000 & 0.0000 $\pm$ 0.0000 & 0.1357 $\pm$ 0.0004 \\
\cmidrule(l){2-7}
  & \multirow{3}{*}{1000} & Gemini 3 Pro & 0.0051 $\pm$ 0.0154 & 0.0039 $\pm$ 0.0116 & 0.0071 $\pm$ 0.0213 & 5.2761 $\pm$ 0.0000 \\
  &   & Opus 4.5 & 0.0094 $\pm$ 0.0187 & 0.0075 $\pm$ 0.0151 & 0.0113 $\pm$ 0.0225 & 3.2374 $\pm$ 0.0026 \\
  &   & GPT-5.2 & 0.0000 $\pm$ 0.0000 & 0.0000 $\pm$ 0.0000 & 0.0000 $\pm$ 0.0000 & 0.7089 $\pm$ 0.0004 \\
\midrule
\multirow{9}{*}{Q18}
  & \multirow{3}{*}{10} & Gemini 3 Pro & 0.5006 $\pm$ 0.0000 & 0.1879 $\pm$ 0.0142 & 1.8557 $\pm$ 0.1625 & 0.1014 $\pm$ 0.0019 \\
  &   & Opus 4.5 & 0.0000 $\pm$ 0.0000 & 0.0000 $\pm$ 0.0000 & 0.0000 $\pm$ 0.0000 & -- \\
  &   & GPT-5.2 & 0.1017 $\pm$ 0.3052 & 0.0517 $\pm$ 0.1551 & 1.0504 $\pm$ 3.1511 & 0.1172 $\pm$ 0.0029 \\
\cmidrule(l){2-7}
  & \multirow{3}{*}{100} & Gemini 3 Pro & 0.2563 $\pm$ 0.0000 & 0.1382 $\pm$ 0.0080 & 0.8541 $\pm$ 0.0561 & 0.1619 $\pm$ 0.0019 \\
  &   & Opus 4.5 & 0.0000 $\pm$ 0.0000 & 0.0000 $\pm$ 0.0000 & 0.0000 $\pm$ 0.0000 & -- \\
  &   & GPT-5.2 & 0.0616 $\pm$ 0.1849 & 0.0389 $\pm$ 0.1166 & 0.5274 $\pm$ 1.5823 & 0.1783 $\pm$ 0.0029 \\
\cmidrule(l){2-7}
  & \multirow{3}{*}{1000} & Gemini 3 Pro & 0.0481 $\pm$ 0.0000 & 0.0414 $\pm$ 0.0008 & 0.0592 $\pm$ 0.0012 & 0.6997 $\pm$ 0.0019 \\
  &   & Opus 4.5 & 0.0000 $\pm$ 0.0000 & 0.0000 $\pm$ 0.0000 & 0.0000 $\pm$ 0.0000 & -- \\
  &   & GPT-5.2 & 0.0187 $\pm$ 0.0562 & 0.0159 $\pm$ 0.0477 & 0.0737 $\pm$ 0.2211 & 0.5334 $\pm$ 0.0029 \\
\midrule
\multirow{9}{*}{Q21}
  & \multirow{3}{*}{10} & Gemini 3 Pro & 0.0000 $\pm$ 0.0000 & 0.0000 $\pm$ 0.0000 & 0.0000 $\pm$ 0.0000 & -- \\
  &   & Opus 4.5 & 0.3922 $\pm$ 0.1961 & 0.1984 $\pm$ 0.0992 & 1.2211 $\pm$ 0.6107 & 0.2233 $\pm$ 0.0003 \\
  &   & GPT-5.2 & 0.0000 $\pm$ 0.0000 & 0.0000 $\pm$ 0.0000 & 0.0000 $\pm$ 0.0000 & 0.4713 $\pm$ 0.0000 \\
\cmidrule(l){2-7}
  & \multirow{3}{*}{100} & Gemini 3 Pro & 0.0000 $\pm$ 0.0000 & 0.0000 $\pm$ 0.0000 & 0.0000 $\pm$ 0.0000 & -- \\
  &   & Opus 4.5 & 0.2300 $\pm$ 0.1150 & 0.1462 $\pm$ 0.0731 & 0.6527 $\pm$ 0.3264 & 0.3002 $\pm$ 0.0003 \\
  &   & GPT-5.2 & 0.0000 $\pm$ 0.0000 & 0.0000 $\pm$ 0.0000 & 0.0000 $\pm$ 0.0000 & 0.7582 $\pm$ 0.0000 \\
\cmidrule(l){2-7}
  & \multirow{3}{*}{1000} & Gemini 3 Pro & 0.0000 $\pm$ 0.0000 & 0.0000 $\pm$ 0.0000 & 0.0000 $\pm$ 0.0000 & -- \\
  &   & Opus 4.5 & 0.0308 $\pm$ 0.0154 & 0.0286 $\pm$ 0.0143 & 0.0241 $\pm$ 0.0121 & 1.5039 $\pm$ 0.0003 \\
  &   & GPT-5.2 & 0.0000 $\pm$ 0.0000 & 0.0000 $\pm$ 0.0000 & 0.0000 $\pm$ 0.0000 & 4.4840 $\pm$ 0.0000 \\
\bottomrule
\end{tabular}
\end{table}

\end{document}